# X-ray Studies of Structure and Defects in Solid $^4$He from 50 mK to Melting

C. A. Burns[1], N. Mulders[2], L. Lurio[3], M. H. W. Chan[4], A. Said[1,5], C. N. Kodituwakku[1,5], and P. M. Platzman[6]

[1] *Department of Physics, Western Michigan University, Kalamazoo, Michigan 49008*
[2] *Department of Physics and Astronomy, University of Delaware, Newark, Delaware 19716,*
[3] *Department of Physics, Northern Illinois University, Dekalb, Illinois 60615*
[4] *Department of Physics, the Pennsylvania State University, University Park, Pennsylvania 16802*
[5] *XOR, Advanced Photon Source, Argonne National Laboratory, Argonne, Illinois 60439*
[6] *Bell Labs, Alcatel-Lucent, Murray Hill, New Jersey 07974*

*Recent measurements have found non-classical rotational inertia (NCRI) in solid $^4$He starting at T ~ 200 mK, leading to speculation that a supersolid state may exist in these materials. Differences in the NCRI fraction due to the growth method and annealing history imply that defects play an important role in the effect. Using x-ray synchrotron radiation, we have studied the nature of the crystals and the properties of the defects in solid $^4$He at temperatures down to 50 mK. Measurements of peak intensities and lattice parameters do not show indications of the supersolid transition. Using growth methods similar to those of groups measuring the NCRI we find that large crystals form. Scanning with a small (down to 10 x 10 $\mu m^2$) beam, we resolve a mosaic structure within these crystals consistent with numerous small angle grain boundaries. The mosaic shows significant shifts over time even at temperatures far from melting. We discuss the relevance of these defects to the NCRI observations.*

Theoretical speculation about supersolidity in $^4$He goes back over 40 years. While BEC will not occur in classical solids[1], solid helium is strongly influenced by quantum effects. Due to its large zero point energy and weak interatomic attraction, $^4$He requires pressures of ~25 bar to solidify. Atoms in the solid are not well localized; vibrations due to the zero point energy can be over 25% of the interatomic spacing[2].

Several mechanisms for bulk supersolid formation have been suggested. Andreev and Lifshitz[3], and Chester[4], argued that a finite concentration of vacancies might exist even at $T = 0$ which could then lead to a supersolid state. Leggett[5] argued that exchange could give rise to supersolidity, although with a rather small supersolid fraction. While a number of experimental searches were initiated,[6] no evidence was found for a supersolid state. Several interesting effects were observed by Goodkind's group,[7] but these results were not consistent with the standard supersolid picture.

The situation changed radically in 2004 when Kim and Chan reported[8] a decrease in the moment of inertia (called NCRI – non-classical rotational inertia) of solid $^4$He in torsional oscillator (TO) experiments near 200 mK. Experiments in porous gold, vycor, and on bulk solid all showed a NCRI of ~1%. NCRI has also been observed in other laboratories[9,10,11,12]. Later work[13] found that the NCRI on a rapidly frozen sample can be as much as 20%, but that this value was greatly reduced by annealing. On the theoretical side, path integral Monte Carlo simulations do not find off diagonal long range order (ODLRO) in a defect-free hcp crystal of $^4$He[14]; for a solid to be superfluid zero-point vacancies or interstitials must be an integral part of the ground state[15], but models seem to indicate that such a state is unstable.[16]

A generally puzzling feature is the high transition temperature. Consider the standard equation for the Bose condensation temperature (BEC) for non-interacting particles

$$T_{BEC} = \left( \frac{n}{2.612} \right)^{2/3} \frac{h^2}{2\pi m k_B} \qquad (1)$$

where $n$ is the mass per volume, $h$ is Planck's constant, $m$ is the mass of the boson, and $k_b$ is Boltzmann's constant. Interactions reduce the BEC condensation temperature and strong enough interactions eliminate the effect entirely. The supersolid's transition temperature is ~10x lower than the superfluid's. For a mass equal to the helium mass, this implies a density for the condensate of ~3% of the density of the superfluid. If vacancies or other defects are causing the transition they are either exceedingly numerous or have a very low effective mass. Anderson[17] has proposed that the observed NCRI effects may be vortex effects at temperatures well above the true supersolid transition temperature.

Pressure driven flow experiments do not show flow[18] in Vycor or 25 μm diameter straight channels, again providing evidence against the simple bulk supersolid model. However, work by M.W. Ray and R. B. Hallock[19] does find evidence for flow through solid $^4He$ in at least some cases. Specific heat measurements[20] show evidence of a peak, near 75 mK, the same temperature of the onset of NCRI in ultrapure $^4He$[21] (with only 1 ppb $^3He$).

The fact that annealing (which should lead to better crystal quality) *reduces* the superfluid fraction implies that defects are likely to play an important role. Adding even a few ppm $^3He$ to the solid *increases* the transition temperature, a surprising feature that may relate to the binding of $^3He$ to dislocations.[22] Recent measurements of the shear modulus at low frequencies and low strains by Day and Beamish[23] find a large increase with a similar dependence on temperature, $^3He$ concentration, amplitude, and annealing as seen in the TO measurements. The authors argue that their data support the idea that there is an important role for dislocations in the effect. In order to understand the transition and the affects of annealing on it, an understanding of the basic structure in the crystal and the defects present is of great importance.

A number of samples of solid helium were grown from standard purity helium at roughly constant pressure near 60 bar (with a melting temperature $T_m \sim 2.6$ K), in a cylindrical Be sample cell (1.9 cm long, inner diameter 3 mm, 1 mm thick Be walls). A copper end-cap was attached to the mixing chamber of the dilution refrigerator while the fill line came in the other end. Thermal anchoring to the cell was provided by a sintered silver piece in the fill line close to the cell. This piece had a heater attached to prevent blockage during crystal growth. The refrigerator was a homemade unit[24] which used a CryoMech PT405 pulse tube (base temperature 2.5 K) for pre-cooling the mixture. This unit eliminates the helium bath and there are no transfers to disturb the measurements. Heat shields were aluminum with aluminized mylar windows to minimize background. The refrigerator allowed tilts of ± 20 degrees along or perpendicular to the x-ray beam, and rotations of ± 40 degrees about a vector pointing along the long (vertical) axis of the cryostat. The base temperature for the refrigerator is 45 mK. A low base temperature is critical, since the TO measurements do not show the maximum NCRI fraction until near 50 mK.

X-ray diffraction measurements were taken at an energy of 22 keV at the Advanced Photon Source (APS) undulator beamline 8-ID-E. Measurements were taken in the vertical scattering plane, where the resolution is the highest. No heating was observed with the beam on. Calculations based on the known thermal conductivity of helium and the x-ray absorption indicated that local heating would be less than 0.01 mK even at 50 mK for our incident flux.

Crystal alignment and some measurements were carried out with a CCD camera. A Cyberstar scintillation detector with detector and sample slits was used for more accurate measurements. Angular accuracy for the CCD was limited by crystal size effects (there was no collimation), while the scintillator's accuracy was determined by the slit settings.

Sample growth occurred at fixed pressure, with a temperature gradient across the length of the sample cell. The loss of the diffuse liquid structure factor ring on the CCD indicated formation of the solid. The samples were not annealed. Several hours were typically required to find a single peak on the CCD. In most cases all orientation scans with the CCD could be consistently indexed as a single crystal (though with some mosaic structure as described later). Due to the experimental geometry, our measurement of crystal size is essentially limited to its 2-*d* projection in a plane perpendicular to the beam. However, in this geometry we found the crystallites to vary from ~1 mm to the limits of our scans (about 10 mm) in the relevant dimensions. Faster freezing created the smaller (~mm) crystals. There was no evidence for any powder, liquid, or amorphous component, although the weak scattering from liquid or amorphous phases means we cannot rule out a small amount of either of these. Molar volumes were determined from the lattice parameters and consistent with the growth conditions.

We searched for evidence of the supersolid transition by studying the lattice parameter and integrated peak intensities through the supersolid transition temperature on a large crystallite. Any bulk transition will involve either the helium atoms or lattice vacancies. Changes in the properties of the lattice are likely to be reflected in the lattice parameter or its slope as a function of temperature. In addition, the average kinetic energy of the atoms should be reduced as the particles enter the condensate. This effect is seen in liquid helium below the superfluid transition temperature[25]. A change in the atomic kinetic energy should be reflected in the zero point motion.

The insert to Fig. 1a shows raw data for a (202) peak from a helium crystal with a molar volume of 18.1 $cm^3$. The width is consistent with the experimental width due to the slits and geometry, and indicates a lower limit on the correlation length of about 1000 Å. Fig 1a shows the peak position for this crystal as a function of temperature. Data were taken both on warming and cooling with no systematic difference between the two. No change is seen at or below the transition to an accuracy of about $3x10^{-5}$. Our data is consistent with earlier neutron work[26,27,28] which had a similar level of accuracy. The line shapes of the peaks show no indication of any alteration in their structure. The spread in the data is much larger than the random error in the peak position. It is consistent with changes due to motion of the sample that result from thermal expansion/contraction of the aluminum refrigerator mount as the temperature in the hutch changes. With our design a 1K change in room temperature would result in a ~25 micron change in the sample height, which changes the region of mosaic we observe a bit, and therefore the intensity. Improved accuracy by a factor of ~20 should be possible using an analyzer crystal before the detector and a more stable mount.

Fig. 1b shows the intensity as a function of temperature for the same reflection. Variations in the intensity are larger than one would predict based on experimental uncertainties and again may be due to the small sample motion mentioned above. The intensity of an x-ray rocking curve is given by the integrated intensity $I_i$ multiplied by the Debye-Waller factor[29]

$$I = I_i \exp[-\tfrac{1}{3}\langle u^2 \rangle G^2] \quad (2)$$

where $\langle u^2 \rangle$ is the mean square deviation of the atoms about their lattice sites, and $G$ is the reciprocal lattice vector. $I_i$ contains a number of experimental and geometric factors, but all are independent of temperature, so any change in intensity indicates a change in $\langle u^2 \rangle$. Averaging the measurements and fitting the points as a function of temperature yields an uncertainty in the intensity of about 1% which corresponds to an uncertainty in $u^2$ of ~0.4%.

Bose condensation would affect the zero point motion in two ways. First, there will be a change (reduction) in the average atomic kinetic energy. Second, the potential between the atoms may be altered. The potential for helium is known to be highly anharmonic and the lattice point for the atoms is not a minimum energy position in the harmonic model. However, the self-consistent harmonic approximation, which creates an effective harmonic model by using a potential which is averaged over the atomic motion of all the particles, has been quite successful[30]. Phonons in solid helium are well defined excitations and not excessively broad, so an effective harmonic model has some validity. We assume the atomic potential is a function of $u$, where $u$ is the deviation in the atom's position from its equilibrium site.

To estimate an upper bound on the kinetic energy we assume that $K \approx \frac{1}{2}ku^2$, where $k$ is an effective spring constant. With this approximation and the assumption that the interatomic potential does not change significantly at the transition, the uncertainty in $K$ is the same as the uncertainty in $u^2$, that is ~ 0.4%. Note that a change in the potential should result in a change in the lattice constant, so the assumption that the potential does not change is consistent with the data in Fig. 1a.

Neutron data[26,27] also finds that $K$ remains the same within error (about 2% for each data point), and the authors use this data to set limits on the condensate fraction. Neutron studies[28] of $\langle u^2 \rangle$ down to 140 mK also failed to see any changes. For $K$ equal to zero for the condensate fraction limits on the condensate fraction $n_0$ are the same as limits on $K$, namely 0.4% in our case. However, the superfluid fraction and the condensate fraction are not necessarily equivalent; superfluid helium exhibits a 100% superfluid fraction when neutron scattering indicates only ~10% condensate fraction. A 1% superfluid fraction with a 0.1% condensate fraction therefore is not unreasonable. *So, our data (and the neutron data) remain insufficiently accurate to rule out a standard form of BEC similar to that seen in the superfluid.*

The observation that annealing can reduce or eliminate the supersolid fraction argues that non-thermodynamic defects play a role in the creation of a supersolid. Observed NCRI may occur due to changes in the grain boundaries or dislocations. An understanding of the defects in the system is therefore of interest. Recent simulations[31] have argued that the grain boundaries may be superfluid in most cases, so observations of their properties is important.

We have carried out measurements with a small x-ray beam at a fixed energy, which allows us to observe small angle grain boundaries. Fig. 2a shows a scan of the crystal taken sitting on a Bragg peak (that is, fixed detector and crystal angles) and moving the sample position along the directions perpendicular to the beam. Data was taken at a temperature of 60 mK with a slit size of 100 x 100 $(\mu m)^2$. There is always signal, but the intensity of the scattering from the crystal varies by factors of 10. Based on scans at several positions, we find that the proper angle for the crystal changes slightly at different locations, that is, there exists significant mosaic. There are long linear discontinuities separating different regions of the crystal which probably correspond to grain boundaries.

Figure 3a shows the effect of rotating the angle that the crystal axis makes with the incident beam. For these measurements we have fairly wide detector slits so we are accepting all of the scattered radiation. We observe a finite number of large mosaic regions in the crystals. By using small (10 x 10 $\mu m^2$) incident slits and scanning sample position we find that the size of these mosaic regions (in the direction perpendicular to the beam) varies from ~ 40 to a few hundred microns. We see well defined features instead of a broad peak since our beam size is comparable with the mosaic size. These measurements are consistent with a tilt from low angle grain boundaries. These can be described as arrays of dislocations. The angle of the tilt is given by $\theta = b/D$, where $b$ is the magnitude of the Burgers vector for the dislocation and $D$ is the approximate distance between the dislocations. Using a tilt angle for the mosaic of 6.5 x $10^{-4}$ rad (see the notation in Fig. 3a) we find that the distance between dislocations is approximately 1500*b*.

Previous synchrotron studies of crystal boundaries in solid $^4$He and solid $^3$He have been carried out by means of x-ray topography.[32] These authors used a white x-ray beam to study the evolution of sub-boundaries between crystal regions. Because of their beam size and film detection scheme, they were sensitive to large angle grain boundaries and could not directly observe the small angle grain boundaries seen here. In $^3$He (but not $^4$He) they were able to observe motion of their grain boundaries. Motion was also observed in bcc $^4$He at higher temperatures using neutron scattering.[33] These authors thought that vibrations were a likely cause of the motion in their case. Because of the limited pressure-temperature range over which the bcc phase survives, the measurements were always close to the melting curve and may not be relevant to $^4$He. Some data was taken on solid $^4$He by the authors, but there was insufficient detector resolution to study changes in $^4$He. Our angular resolution is much higher, and we observe changes in peak positions on angular scales which would not be visible with their setup.

We observe that the mosaic regions are capable of moving with respect to each other at higher temperatures. Fig. 3b) shows the time evolution of a particular mosaic structure at $T$ = 1.75 K, well below $T_m$ ~ 2.6 K. Since the regions move with respect to each other, it is clear we cannot have an experimental artifact or rotation of the entire crystal. In general you would not expect a region with an arbitrary shape to be able to move with respect to other regions with random shapes unless the grain boundary was also able to move. In fact, as has been pointed out,[33] a mobile grain boundary moving through a crystal region can cause such a change in scattering angle. The changes in the shape of the peak also provide evidence for motion of the defects/grain boundary through the crystal. The coherent nature of the change (the fact that the peak does not shift randomly back and forth but continues to shift in the same direction) argues for motion driven by a gradient such as stress or pressure. The motion of the peak to the left of the main peak (indicated by the arrow in Fig. 3b corresponds to an angular velocity for the crystallite of ~1.6 x $10^{-7}$ rad/s. While there is a spread in the speed of motion for different crystallites at a fixed temperature, the overall rate increases with *T*. This increase with temperature argues for classical motion over some sort of energy barrier. Fig. 4 shows the evolution of the peak at different temperatures. In this figure, each curve takes ~ 5 minutes to acquire and there is only a short gap between scans for the motors to restart. There are a number of noteworthy features.

First, the evolution of the peaks clearly increases with temperature. In fact, at the highest temperature shown (which is still far from the melting temperature of 2.6 K), there is a strong redistribution of spectral weight with temperature. The crystals are dynamic entities at the higher temperatures. Second, the crystals are likely to have significant strains or pressure gradients frozen in; simple thermal fluctuations of the peaks would cause a random evolution of the peaks. Peaks observed in the neutron studies[33] moved randomly, while we see evolution of the peaks in a certain direction. Third, there is a large change in the shape of the peaks between different temperatures, that is, the act of changing the temperature results in a redistribution of the peaks. The time to change the temperature by 100 mK was on the order of 5 minutes. Stress induced by changing the temperature therefore can easily alter the crystal structure, and slow temperature changes may be needed to avoid stress in the crystals.

The integrated intensity does not remain the same between the different measurements, mainly due to the migration of structure out of the scan range. We note that on some occasions in other crystals we have observed significant alignment changes for the whole crystal (~0.5°) at temperatures within about 0.2 K of melting.

How do our measurements relate to the NCRI? First. the number of grain boundaries appears to be too small to directly lead to a 1% supersolid fraction. However, this does not rule out the possibility that they play a role. If grain boundaries became supersolid, it may be possible for a region of normal crystal to rotate freely over a small range of angles. This is similar in some aspects to the model of Prokof'ev and Svistunov[15] who postulated superfluid boundaries between small crystallites, but here mosaic regions would be separated by supersolid grain boundaries. In such a case, a very small volume supersolid fraction can lead to relatively large effects on the TO measurements. Also, it is quite likely that superflow in such a system would be hindered, as observed. However, this idea cannot explain the data in Vycor and porous gold, which show a similar supersolid fraction compared to the bulk.

Another possibility is that the supersolid effect only exists in a limited subset of these mosaic regions. This is consistent with the fairly broad transition temperature, which may be due to inhomogeneous regions with a distribution of transition temperatures. It is also consistent with the fact that while the NCRIf appears to change, the starting transition temperature (at fixed $^3$He concentration) appears to stay the same. Increasing the number of regions that should the effect would increase the fraction of the volume that had NCRI but not the transition temperature.

The changes in the peak structure with time or due to small changes in temperature indicate that these crystals are easy to strain. An understanding of the NCRI effect may therefore require explicit treatment of the effects of strain on the crystal properties.

In conclusion, we have used x-ray synchrotron radiation to study the properties of hcp solid $^4$He crystals. The high angular resolution coupled with the ability to use small beams makes x-rays a useful probe to study these systems. Consistent with other workers, we find no changes in the lattice constant or peak intensity which would indicate a bulk transition in the solid, although the current accuracies are insufficient to completely rule out the possibility of bulk supersolid behavior. Solidification seems to result in large but imperfect crystal formation. We have used a narrow x-ray beam to study the defects in $^4$He crystals, and find that crystals contain a microstructure of mosaic regions consistent with small angle grain boundaries. These regions are capable of moving with respect to each other, but the motion is only observed at temperatures much higher than the transition temperature. The motion is not random, and is probably due to the relaxation of strains or pressure gradients in the crystal. Changing the temperature also alters the mosaic structure.

Acknowledgements: CAB acknowledges support for this project from DOE under grant DE-FG01-05ER05-02 and MHWC acknowledges support from NSF grant DMR 020701 and DMR 0706339. We thank the staff at Sector 8 of the APS for the assistance with the measurements. Use of the Advanced Photon Source at Argonne National Laboratory was supported by the U. S. Department of Energy, Office of Science, Office of Basic Energy Sciences, under Contract No. DE-AC02-06CH11357.

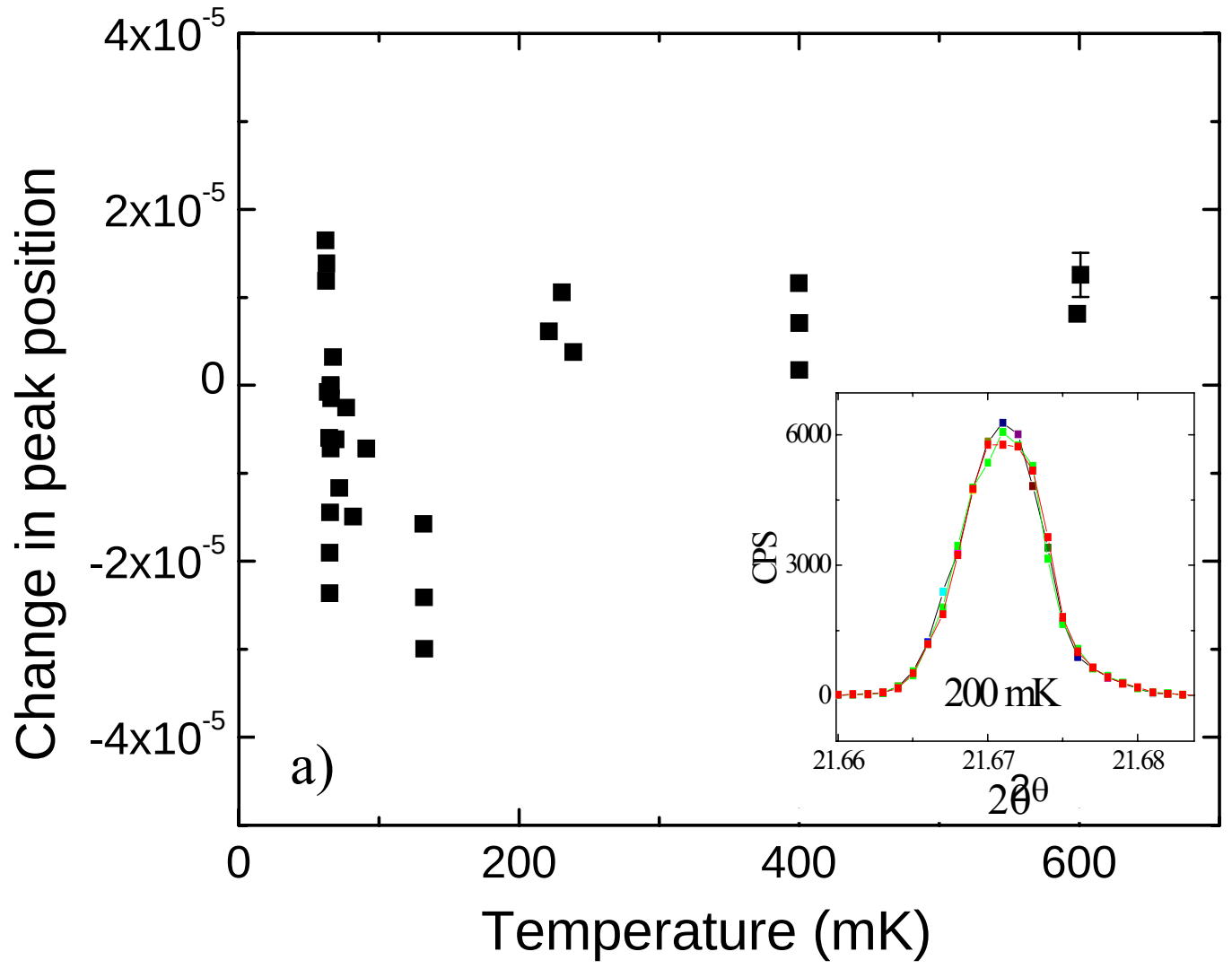

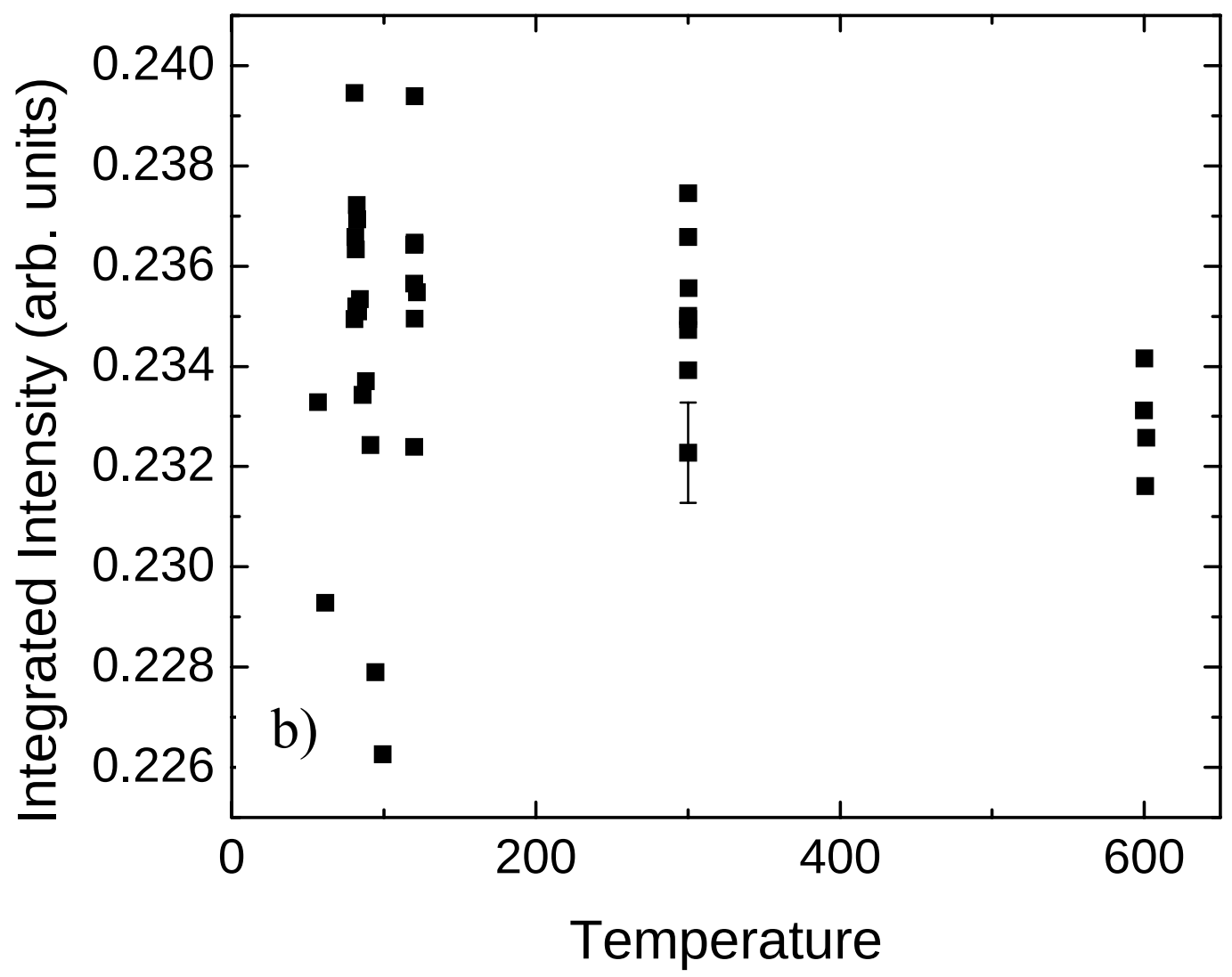

FIG. 1 a) Change in lattice position as a function of temperature for a (202) reflection. Insert shows three separate scans of peak at the same temperature. b) Integrated intensity for the (202). Neither data set shows any changes at or below the start of the transition temperature (~200 mK).

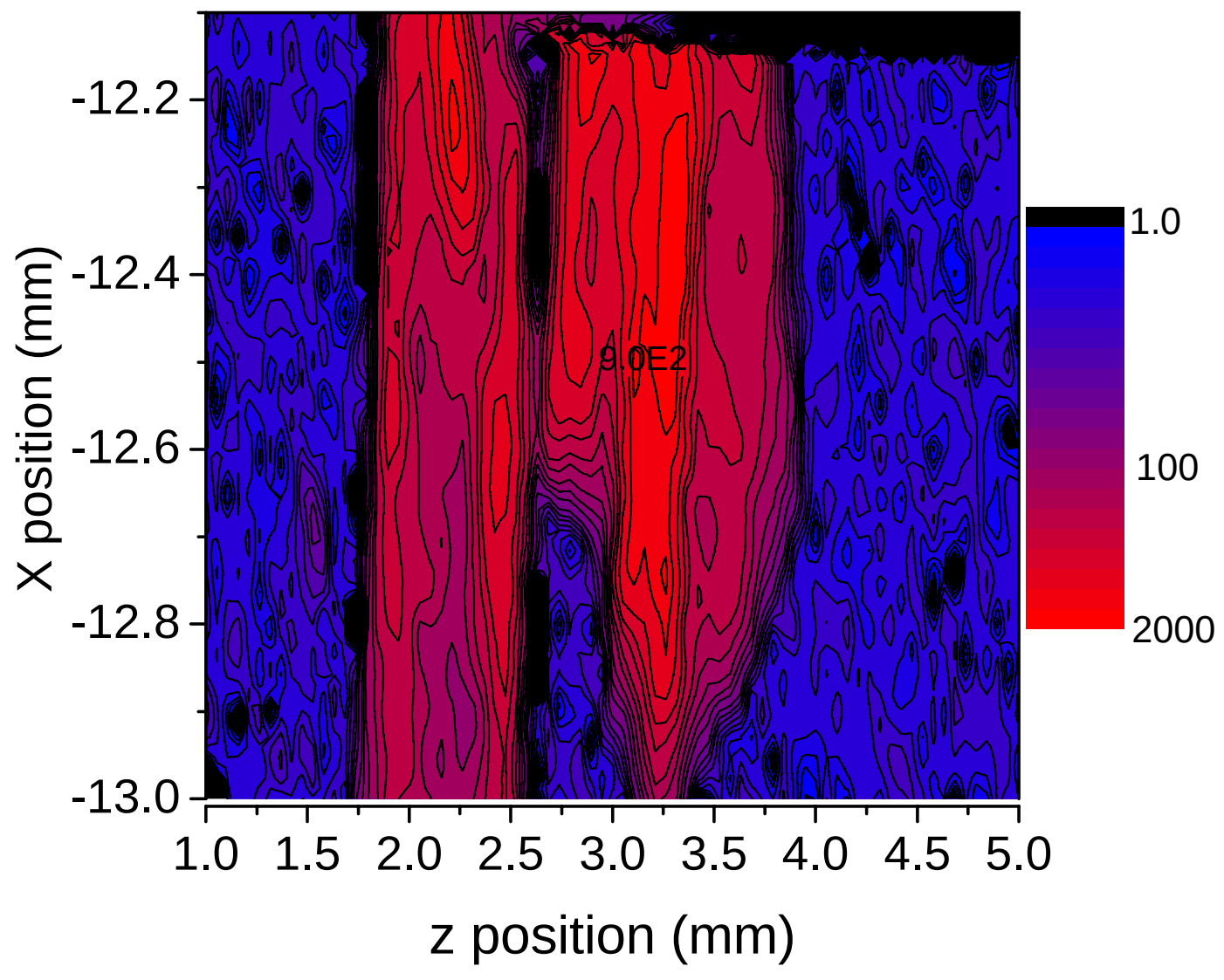

Fig. 2 (color online) Bragg intensity as a function of physical position on the sample at fixed sample angle for the (202) reflection. Linear resolution was 100 μm in each of the *x* and *z* directions (perpendicular to the beam). This figure shows regions of the crystal that have the same alignment to the beam. The regions with weaker scattering are slightly misaligned. Regions with different orientations are separated by (linear resolution limited) sections that are long and straight – that is, grain boundaries.

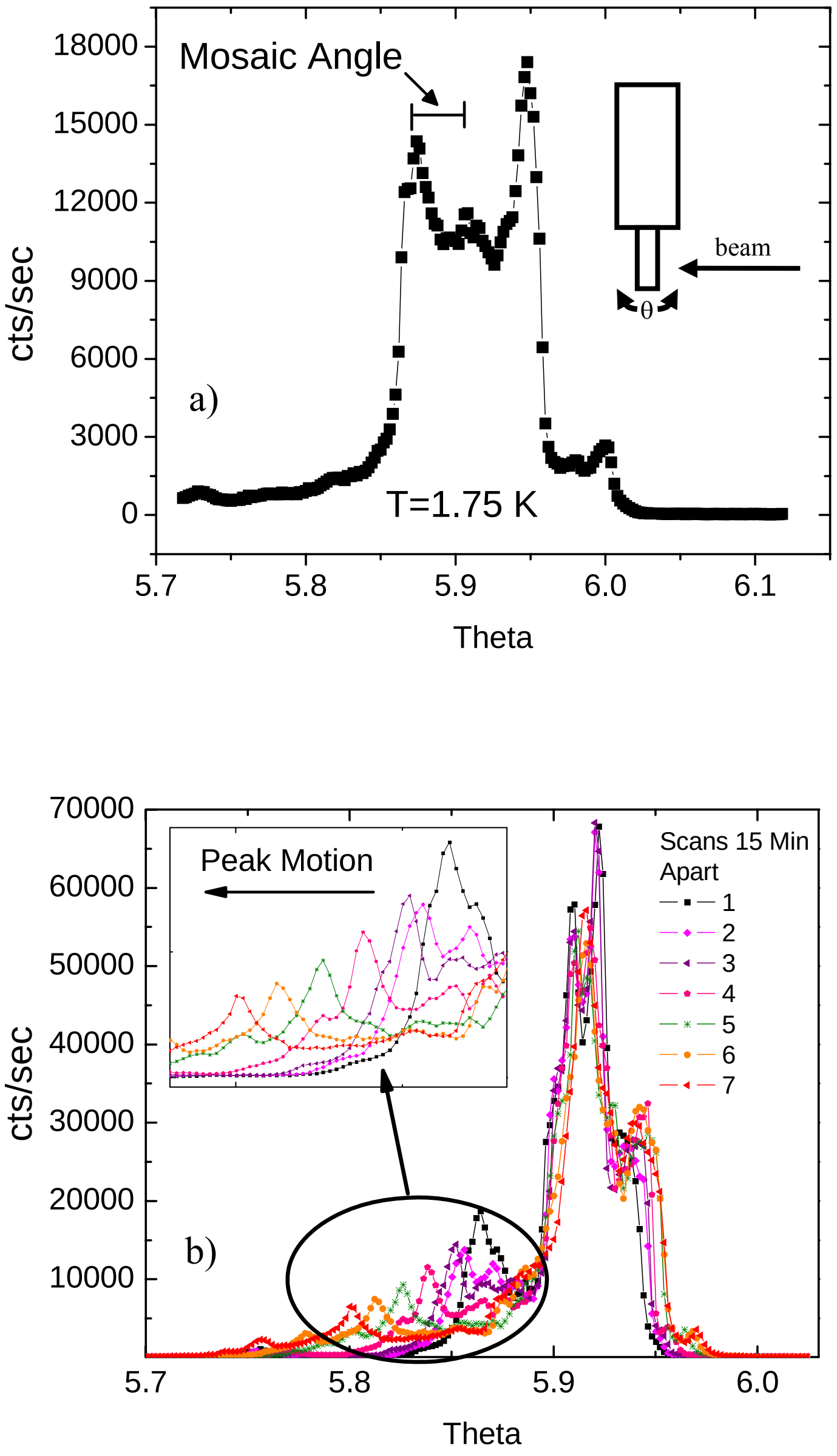

Fig. 3 a) Intensity vs. crystal angle theta for (101) reflection. Note that the crystal shows a strong mosaic structure. However, there is not a continuous distribution of peaks but a small number of distinguishable crystal regions with sizes on the order of tens to hundreds of microns.

b) (color online) Intensity vs. crystal angle at $T = 1.75$ K ($T_m \sim 2.6$ K). Scans were taken fifteen minutes apart and show the variation with time of the mosaic structure. These changes indicate that the crystal structure does not become "frozen in" even well below the melting temperature.

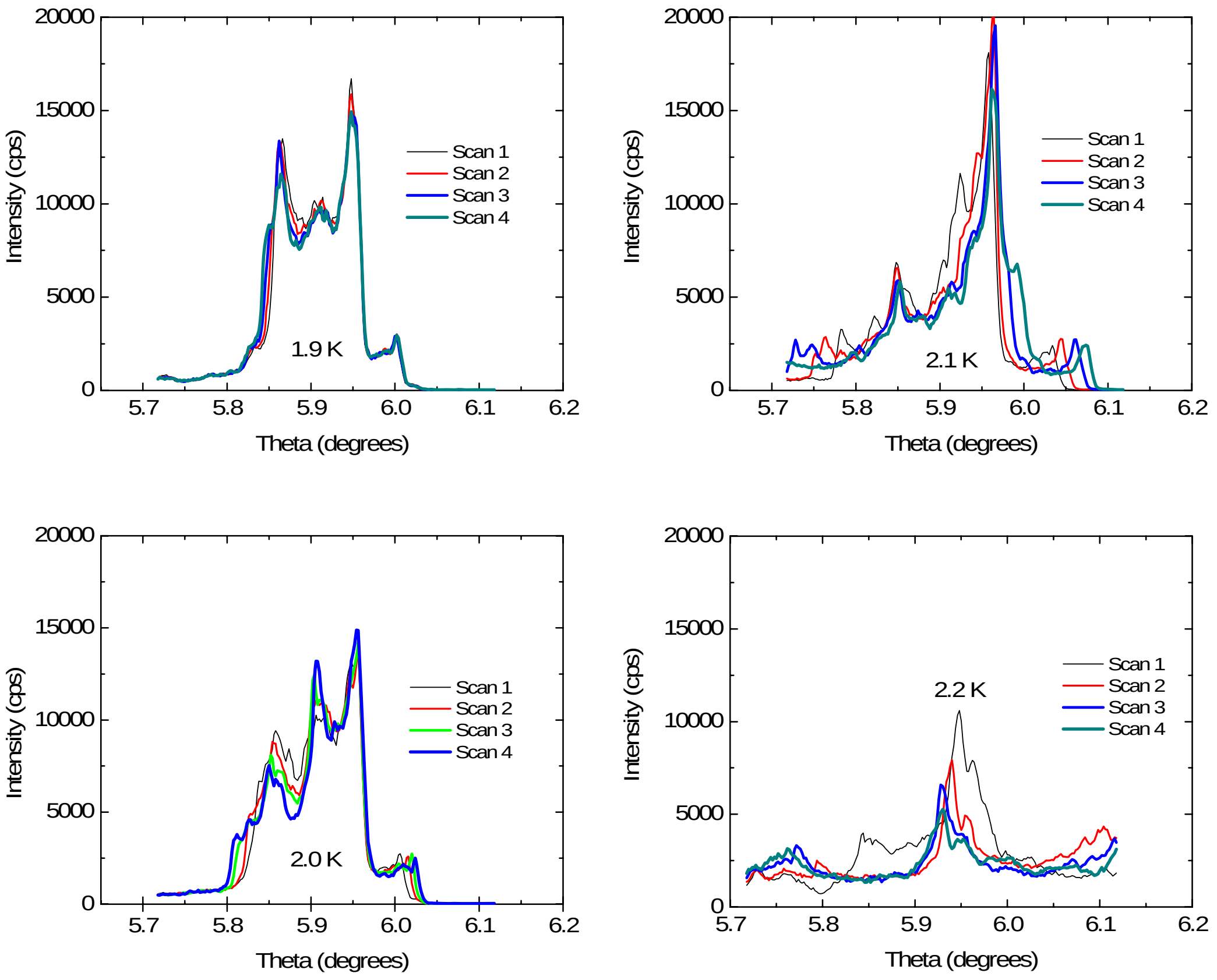

Fig. 4 (color online). Temperature dependence of the mosaic structure of the (101). The four scans at a given temperature are taken at about 5 minute intervals. Scans later in time are plotted with thicker lines. At higher temperatures, there is a clear alteration of the spectrum with time. Note that there are significant changes *between* scans, that is altering the temperature alters the mosaic.

---

[1] O. Penrose and L. Onsager, Phys. Rev. 104, 576 (1956).

[2] C. A. Burns and E. D. Isaacs, Phys. Rev. B, 55, 5767 (1997).

[3] A. F. Andreev and I. M. Lifshitz, Sov. Phys. JETP 29. 1107 (1969).

[4] G. V. Chester, Phys. Rev. A 2, 256 (1970).

[5] A. J. Leggett, Phys. Rev. Lett. 25, 1542 (1970).

[6] M. W. Meisel, Physica B, 178, 121 (1992).

[7] J. M. Goodkind, Phys. Rev. Lett. 89, 095301 (2002).

[8] E. Kim, M. H. Chan, Nature 427, 225 (2004).

[9] A. S. Rittner and J. D. Reppy, Phys. Rev. Lett. 97, 165301 (2006).

[10] Y. Aoki, J.C. Graves, and H. Kojima, Phys. Rev. Lett. 99, 015301 (2007).

[11] M., S. Kondo, Takada, Y. Shibayama, and K. Shirahama, J. Low Temp. Phys. 148, 695 (2007).